\documentstyle[12pt]{article}
\newcounter{saveeqn}
\newcounter{App} 

\def\aa{{\cal A}}
\def\be{\begin{equation}}
\def\ee{\end{equation}}
\def\bq{\begin{eqnarray}}
\def\eq{\end{eqnarray}}

\def\g{\gamma}
\def\ra{\rightarrow}

\def\n{\nonumber}

\begin{document}

\begin{flushright}
CEBAF-TH-94-23\\
HEP-LAT/9412033
\end{flushright}
\vspace{1cm}
\begin{center}
{\bf Massless Fermions on the Lattice.}
\end{center}
\begin{center}{V.M. Belyaev$^*$}
\end{center}
 \begin{center} 
 Continuous Electron Beam Accelerator Facility  
12000 Jefferson Ave, Newport News, Virginia 23606, USA
\end{center}
\vspace{1cm}
\begin{abstract}
We consider a nonlocal lattice action for fermions   fermion doubling
in lattice theories.

It is shown, that it is possible to 
avoid the fermionic doubling in the case of free fermions,
 but this approach does not reproduce
results for the effective action for  gauge fields in the continuum theory,
because the  high frequency fermion modes have a strong dependence on
the gauge field.
\end{abstract}
\vspace{1cm}
\flushbottom{$^*\overline{On\;  leave\;  of\;  absence }
\;from\;  ITEP,\;  117259\;  Moscow,\; Russia.$}

\newpage

\section{Introduction}
Nowadays, lattice techniques have become an  important method for the 
study of non-perturbative quantum
field theory. It is widely used for calculations of hadron spectrum and for 
understanding the structure
of nonperturbative effects in quantum field theory \cite{review}. 
However, there is a serious problem to avoid
the fermion doubling and to implement the axial anomaly into the lattice field 
theory. 
The most popular way to resolve the problem was suggested by
Wilson
 \cite{wilson} and in ref.\cite{anom} it was shown, that in the continuum
 limit, the lattice theory with Wilson fermions reproduces axial anomaly. 
The Wilson action contains a term which breaks chiral invariance by mass-like
term, with a mass parameter of the order of inverse size of a lattice size and, 
due to interactions,
this term is renormalized and    fine tuning is then required. The 
Kogut-Susskind action \cite{kogut} reduces the number of
fermions but does not reproduce the axial anomaly. 
Also, there are attempts to resolve the problem  by a gauge-violating
Majorana-type Wilson mass \cite{pryor}. 

Recently, there has been a considerable theoretical activity on the use of 
surface states as a basis 
for a theory of chiral lattice fermions \cite{kaplan} where our four dimensional 
world
is considered as an interface in a five dimensional underlying space 
\cite{altes93}.
 Under some conditions, low energy fermionic states are bound to this wall.
For this low energy states, we have an effective chiral theory on the interface. 
A Hamiltonian
approach for this formulation of the chiral theory was considered 
in ref.\cite{creutz}.

In  this paper, we   study nonlocal gauge invariant 
action for  massless fermions without fermion doubling.

\section{The doubling problem and nonlocal lattice action for  fermions.}
Let us consider a general one-dimensional lattice action for a free massless 
 fermion field:
\bq
S=\sum_{n,m}\bar{\psi}_m A_{m,n}\psi_n
\label{1}
\eq
where
\bq
 A_{m,n}=-A_{n,m}=A_{m-n}\;\;\; A(0)=0
\label{2}
\eq

In this paper, we consider the case of antiperiodic boundary conditions,
which  are imposed on the
fermion field for convenience only. 
The case of periodical boundary conditions could be considered as well;
nothing would be changed except minor technical details.

Let us consider the lattice with even numbers of elements, 
\bq
-N\leq n\leq N-1,
\label{2a}
\eq
we conveniently  rewrite the action (\ref{1}) in the following form
\bq
S=\sum_{n,m}\bar{\psi}_{m+\frac12} A_{m-n}\psi_{n+\frac12}
\label{3}
\eq
where $2N$ is the size of the  lattice. 

To diagonalize the action (\ref{1}), we use Fourier transformation:
\bq
\psi_{m+\frac12}=\sqrt{\frac1{2N}}\sum_{k=-N}^{N-1}e^{i\frac{\pi (k+\frac12)( 
m+\frac12)}{N}}\Psi_{k+\frac12}
\nonumber
\\
\bar{\psi}_{m+\frac12}=\sqrt{\frac1{2N}}\sum_{k=-N}^{N+-1}
e^{-i\frac{\pi (k+\frac12) (m+\frac12)}{N}}\bar{\Psi}_{k+\frac12}
\label{4}
\eq
In the new variables, the action (\ref{1}) has the form:
\bq
S=\frac1{2N}\sum_{m,n,k,l}\bar{\Psi}_{k+\frac12}e^{-i\frac{\pi 
(k+\frac12) (m+\frac12)}{N}}A(m-n)e^{i\frac{\pi (l+\frac12) 
(m+\frac12)}{N}}\Psi_{l+\frac12}
\label{5}
\eq
After summation over $m$ for fixed value of $(m-n)$, we obtain:
\bq
S^{fer.}&=&\sum_{k,n=-N}^{N-1}\bar{\Psi}_{k+\frac12} e^{-i\frac{\pi 
(k+\frac12)(n+\frac12)}{N}}A(n)\Psi_{k+\frac12}
\nonumber
\\
&=&-i\sum_{k=-N}^{N-1}\bar{\Psi}_{k+\frac12} \left(2\sum_{n=0}^{N-1}A(n)
\sin (\frac{\pi (k+\frac12)n}{N})\right)\Psi_{k+\frac12}
\nonumber
\\
&=&\sum_{k=-N}^{N+1}\bar{\Psi}_{k+\frac12} B(k+\frac12)\Psi_{k+\frac12}
\label{6}
\eq
where
\bq
B(k)=-i2\sum_{n=0}^{N-1}A(n)\sin (\frac{\pi (k+\frac12)n}{N}).
\label{7}
\eq

In the case of the standard lattice action for fermions
\bq
A(n)=i\frac{\delta^{n1}}{2a},
\nonumber
\\
\omega_{k+\frac12}=\frac{\pi (k+\frac12)}{Na},
\label{8}
\eq
where $a$ is the lattice spacing and $\omega$ is the fermion momentum.
 Then the action
has the well known form:
\bq
S=\sum_k \bar{\Psi}_{k+\frac12}\frac{\sin 
(\omega_{k+\frac12} a)}{a}\Psi_{k+\frac12}
\label{9}
\eq
and for $\omega_{N-\frac12}=\frac{\pi (N-\frac12)}{Na}$, in the limit 
$N\ra\infty$, we obtain an
additional pole in the fermion propagator.

Now, let us consider $B(k)$  for $k=N-\frac12$ in the general nonlocal case:
\bq
B(N-\frac12)&=&2\sum_{n=0}^{N-1}A(n)\sin (\frac{\pi (N-\frac12)n}{N})
\nonumber
\\
&=&2\sum_{n=0}^{N-1}A(n)\sin (\pi n-\frac{\pi n}{2N})
\nonumber
\\
&=&2\sum_{n=0}^{N-1}A(n)(-1)^{n+1} \sin (\frac{\pi n}{2N})
\label{10}
\eq
It is clear,  that if we choose that as $n\ra\infty$
\bq
A(n)\sim  \frac{(-1)^{n+1}}n,
\label{11}
\eq
 the second pole in the fermion propagator will be absent 
in the continuum limit.

It is convenient to choose $A(n)$ in the following form
\bq
A(m-n)=\sum_{\alpha=-N}^{N-1}<E_{m+\frac12}e_{\alpha+\frac12}>
\omega_{\alpha+\frac12}<e_{-(\alpha+\frac12)}E_{n+\frac12}>
\label{12}
\eq
where we use the following notations:
\bq
e_{\alpha+\frac12}(x)=\sqrt{\frac1{L}}e^{i\omega_(\alpha+\frac12) x}
\nonumber
\\
E_{m+\frac12}(x)=\sqrt{\frac1a}\Theta\left(x-ma\right)
\Theta\left( (m+1)a-x\right)
\nonumber
\\
<f>=\int_{-L/2}^{L/2}f(x)dx
\label{13}
\eq
Then, in the limit $N\ra\infty$, we obtain:
\bq
A(n)&=&i\frac1a\int_0^a\frac{dx}{\pi}\sin (xn)\frac{2(1-\cos (x))}{x}
\label{14}
\\
B(k+\frac12)&=&\omega_{k+\frac12}\left(\frac{2(1-\cos (\omega_{k+\frac12} 
a))}{\omega_{k+\frac12}^2 a^2}\right)
\label{15}
\eq
From eq.(\ref{15}), we see that the second pole of the fermion propagator is
absent. Notice, that asymptotically  $A(n)$ in eq.(\ref{14}) has  the form
(\ref{11}). 

Also, one may  use the simplest choice for $A(m-n)$:
\bq
A(n)=\frac{(-1)^{n+1}}{an}\;\;\;\; n\neq 0
\label{16}
\eq
In this case, the second pole in the propagator of a free fermion 
will be absent in the continuum limit as well. The case
$A(n)\sim(-1)^n$ was considered in \cite{zenkin}.

Thus, we have constructed the  action for a free massless fermion field with
a single fermion.  Still there is a question: does the theory have a correct 
limit at $N\ra\infty$ in the presence of gauge fields.

To study this question, let us consider the lattice fermion in the presence
of an external gauge field.
In the case of $D=1$, $U(1)$ gauge theory with one
fermion, the action has the following form:
\bq
S=\sum_{m,n=-N}^{N-1}\bar{\psi}_{m+\frac12} V^{(m+\frac12)
(n+\frac12)}A(m-n)\psi_{n+\frac12},
\label{17}
\eq
where we introduce the following matrices on the links of the lattice:
\bq
U^{m+\frac12}={\cal P}\exp 
\left(ig\int_{x=(m-\frac12)a}^{x=(m+\frac12)a}\aa(x)dx\right)
\label{18}
\eq
and
\bq
V^{(m+\frac12)(n+\frac12)}&=&U^{m+1/2}U^{m-3/2}...U^{n+1/2}\n
\\
for\;\;\;\; 0
<|m-n|_{mod(2N)}< N
\n
\eq
\bq
V^{(m+\frac12)(n+\frac12)}&=&U^{\dagger (m-1/2)}U^{\dagger (m+1/2)}...U^{\dagger
 (n-1/2)}\n
\\
 for\;\;\;\; -N<|m-n|_{mod(2N)}< 0
\label{19}
\eq

${\cal P}$ denotes a path-ordered product, 
\bq
|m|_{mod(2N)}=m-(2N)j\;\;\; at\;\; -N\leq (m-(2N)j)\leq N-1
\n
\\
j=0,\pm 1,\pm 2 ...
\label{20}
\eq
Here and below we impose periodic boundary conditions for the gauge field $\aa$.
It is known, that we can   gauge away contributions of  nonconstant
components of the gauge field $\aa^n$, where
$
\aa(x)=\sum_n \aa^n e^{i\omega_n x}
$.
But in general case,  it is not possible to remove a constant gauge field
because there are  gauge transformations which violate the boundary conditions:
\bq
\aa\ra\aa-\frac1{g}\partial_x\alpha(x)\n \\
\alpha(x)=g\aa x\n \\
\psi_{n+1/2}\ra e^{i\alpha(x=a(n+1/2))}\psi_{n+1/2}
\label{21}
\eq
and keeps antiperiodic boundary conditions only when
\bq
\alpha(x)=(2\pi/L)nx, \;\;\;n=\pm 1,\pm 2,...
\label{22}
\eq
The presence of the constant gauge field corresponds to the shift
of momentum in eq.(\ref{7}):
\bq
B(k+1/2)\ra B(k+1/2+y/2)
\label{23}
\eq
where $y=\frac{g\aa L}\pi$.

Notice that the highest fermionic modes ($y\sim N$) have a strong dependence on
the gauge field. It means that the contributions of the high frequency modes
in physical observables
(which feel the lattice structure) will not die  in the limit
$N\ra\infty$.

Let us check this statement in the simplest case of the 2-dimensional
Schwinger model with one Dirac fermion. 
In the case of continuous imaginary time and a lattice in space,
the action has the following form:
\bq
S=\int dx\sum_{m,n=-N}^{N-1}\left(\bar{\psi}_{m+1/2}(x)iD_x\g_1\delta^{mn}
{\psi}_{n+1/2}(x)\right.\n
\\
\left.+\bar{\psi}_{m+1/2}(x) V^{(m+\frac12)\g_2
(n+\frac12)}A(m-n)\psi_{n+1/2}(x)\right.
\label{24}
\eq
where $D_x$ is the covariant derivative:
\bq
iD_x\psi_{n+1/2}(x)=(i\partial_x-gA^{n+1/2}_x(x))\psi_{n+1/2}(x)
\label{25}
\eq
It is easy to  calculate the effective action for a constant gauge field
$\aa$. According to the continuous theory, the potential has the following form:
\bq
W(y)-W(0)=N_f\frac{\pi}{2L^2}(|y|^2_{mod(2)})
\label{26}
\eq
where $|y|_{mod(2)}=y+2j$ if $|y+2j|<1$ and $j=0,\pm 1,\pm 2,...$, $N_f$ is the 
number of Dirac fermions.
In the case of the lattice theory we have:
\bq
W(y)-W(0)=-\frac{1}{2L}\int_{-\infty}^{\infty}d\omega\sum_{k=-N}^{N-1}
ln\left(\frac{\omega^2+B^2(k+1/2+y/2)}{\omega^2+B^2(k+1/2)}\right)\n
\\
=-\frac1{2L^2}\sum_{k=-N}^{N-1}\left(|B(k+1/2+y/2)|-
B(k+1/2)|\right)
\label{27}
\eq
We considered the cases of standard  (\ref{8})
and  nonlocal (\ref{16}) fermion actions.
The results obtained for $N=10$ are shown in Fig.1,
where solid lines corresponds to the standard lattice action with
one Dirac fermion  and to  eq.(\ref{26}) for $N_f=2$ (two curves  practically
coincide with each other), and the dashed line corresponds to the case of the
nonlocal lattice action for
 one Dirac fermion field.
 Comparing our results with
eq.(\ref{26}), we see that the standard fermionic action corresponds to
the case of two Dirac fermions in the continuum limit. At the same time
we see that the lattice with nonlocal action for fermions has a strong
dependence on the lattice size and does not tend to any continuous
theory with a finite number of fermions. 
Thus, we see that the lattice effects do not disappear in
the case of nonlocal action. It is possible, of course, to discuss
a possibility to kill this strong dependence by choosing a special
form for $A(m)$, but in this case we need to make a fine tuning and
it is not clear, whether these lattice artifacts disappear in the
other physical values in the limit of $N\ra\infty$.

Thus, we have to conclude that the attempt to resolve the fermion doubling
problem by means of nonlocal action in the case of gauge theory has
 serious difficulties to obtain a correct continuum limit.

\section{Discussions}
In this paper, we considered nonlocal fermion action to solve the problem 
of the fermion doubling. 

It was found that it is possible to construct a
nonlocal lattice action for free fermions without doubling, but
the theory has no correct continuum limit in the presence of a gauge 
field. This problem appears because of the strong couplings of high
frequency modes which feel lattice structure.

\section{Acknowledgements}
I  am grateful to the participants of  the theoretical seminar at CEBAF for 
useful comments on the first part of the paper and
thank J. Goity for very fruitful discussions.

\end{document}